\documentclass[11pt,twoside]{article}
\usepackage{asp2010}

\resetcounters

\begin{document}

\title{Old novae and the SW Sex phenomenon}
\author{Linda~Schmidtobreick$^1$ and Claus~Tappert$^2$ 
\affil{$^1$European Southern Observatory, Casilla 19001, Santiago 19, Chile}
\affil{$^2$Departamento de F\'\i sica y Astronom\'\i a, Universidad de Valpara\'\i so, Avda.\ Gran Breta\~na 1112, Valpara\'\i so, Chile}}

\begin{abstract}
From a large observing campaign, we found that nearly all
non- or weakly magnetic cataclysmic variables (CVs) in the 
orbital period range between 2.8 and 4 hours are of SW Sex 
type and as such experience very high mass transfer rates. 
The exceptions seem to be some old novae that have periods
around 3.5\,h. Their spectra do not show the typical
SW Sex characteristics but rather resemble those of 
dwarf novae with low mass transfer rates. 

The presence of old
novae in this period range of SW\,Sex stars that do not follow 
the trend but show instead rather low mass transfer rates
is interpreted as evidence for an
effect of the nova eruption on the mass transfer rate of the 
underlying CV similar to the hibernation scenario. 
\end{abstract}

%\section{Introduction}
%On first sight, the two fields of SW Sextantis stars and old novae do
%not seem to have much in common. Of course, the systems of both groups 
%are cataclysmic variables (CVs) and some old novae also belong to the 
%group of SW\,Sex
%stars \citep[e.g. the old nova RR\,Pic,][]{schmidtobreicketal03-1}. Still,
%the connections are weak. We therefore start by introducing the two
%fields separately before combining the results of our research in both
%fields and
%using them as evidence for the validity of the hibernation model.

\section{SW Sextantis stars}
The sub-class of SW Sextantis stars 
were originally defined by \cite{thorstensenetal91-1} as eclipsing 
nova-like stars which however show single-peak emission lines with high
velocity line wings, 
strong He\,II emission but no polarisation, and transient absorption
features in the emission lines at an orbital phase of $\phi = 0.5$.
A distinctive feature that was later attributed to SW\, Sex stars is the
orbital phase offset of 0.2 cycles of the radial velocity curves with 
respect to the
photometric ephemeris. In general, SW\,Sex stars are considered
novalikes with an extremely high
mass transfer rate  \citep[see e.g.][]{rodriguez-giletal07-1}.
This idea is supported by the temperature of the white dwarfs in these systems
which exceed the the expected value for accretion governed by an angular
momentum loss from standard magnetic braking \citep{Town+09}.

At the beginning, few SW\,Sex stars were known and they were considered
as strange objects with unusual behaviour. However,
\citet{gaensicke05-1} compared the results of various CV surveys and 
showed that SW\,Sex stars are common in the orbital
period range between 2.8 and 4\,h which is right at the upper edge
of the period gap. He found that
in particular all the eclipsing nova-like systems in this period range
belong to the sub-class of SW\,Sex stars.

Since being an eclipsing system is no
intrinsic physical property of the star but rather
depends on the angle under which the binary is
seen, it is entirely plausible that all
non- or weakly-magnetic CVs just above the
period gap are physically similar to the SW Sex stars, 
and in particular experience a very
high mass transfer rate. We have thus conducted a survey
to test this idea by performing time-series spectroscopy for
non-eclipsing CVs in the 2.8--4\,h period range 
\citep[Schmidtobreick et al. in preparation]{2007MNRAS.374.1359R}. 
We checked for the presence of defining SW Sex characteristics
like broad line wings with large-amplitude radial
velocity variations, single-peaked line profiles
with phase-dependent central absorption, and
phase lags between the radial velocity modulation
in the line cores and wings. Our main result is that
indeed the majority of the observed
CVs in the 2.8--4\,h period range are of SW\,Sex type and can thus be
considered as high mass transfer systems. This suggests 
that SW\,Sex stars actually represent a stage in the secular evolution
of CVs and that CVs reaching the 2.8--4\,h period range will then share
the SW\,Sex characteristics (Schmidtobreick et al.\ in preparation).

\section{Old novae with low mass transfer rates}
\begin{figure}
\resizebox{!}{8.8cm}{\includegraphics{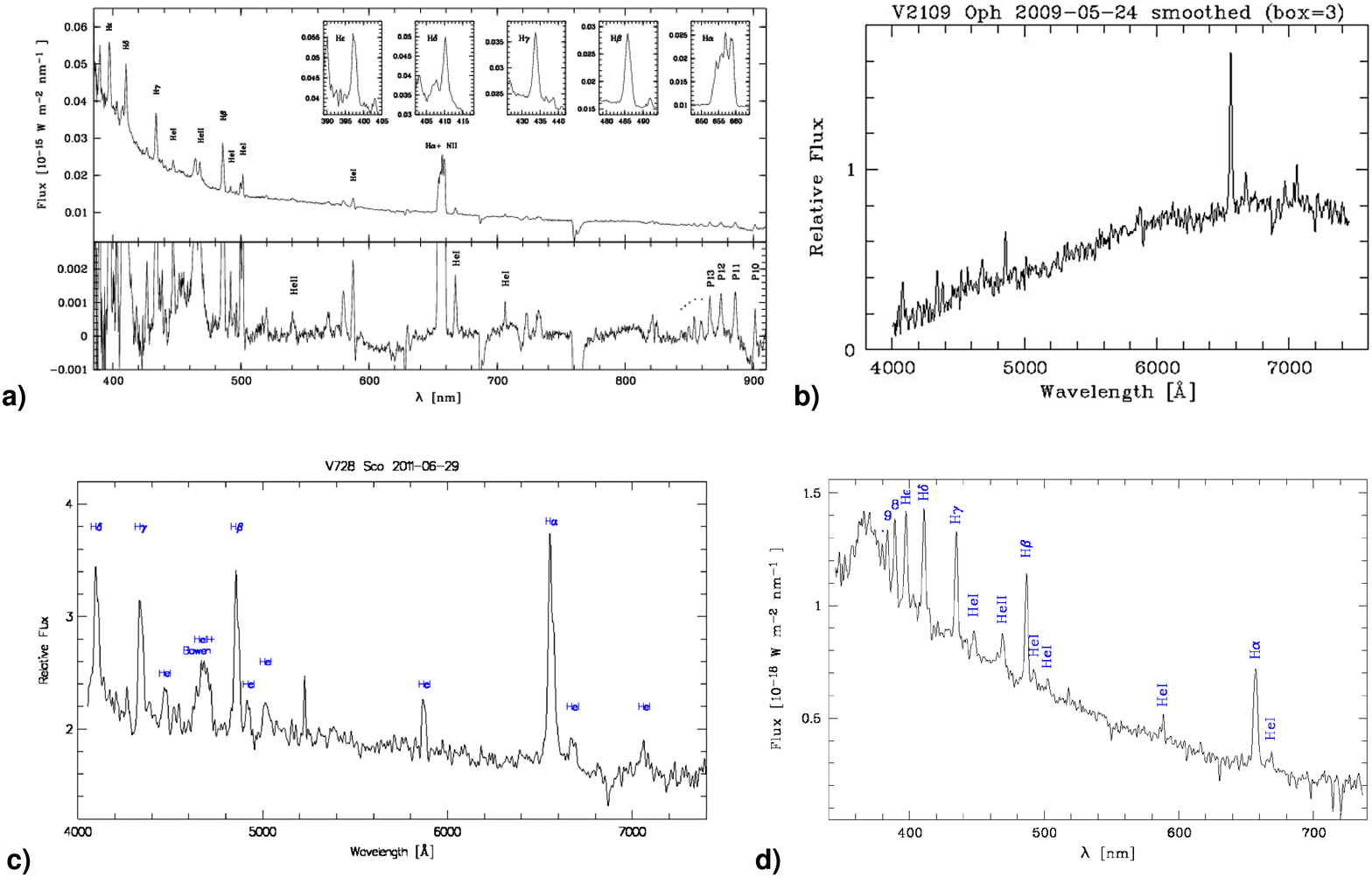}}
 \caption{\label{spectra}
Optical spectra of the four low mass system candidates among
our old nova sample: a) V842\,Cen, b) V2109\, Oph, c) V728\,Sco, d) XX\,Tau.
The plots have been taken from \cite{schmidtobreicketal05-1} (a,d) and 
\cite{tappertetal12-1} (b,c).
}
\end{figure}

Nova eruptions can occur for every CV
assuming that the mass-transfer rate is high enough to
accumulate the necessary amount of material on the surface
of the white dwarf \citep{ibenetal92-1}. 
As the effects of the nova eruption disappear, the behaviour of the
CV depends once again on its properties like orbital period, 
mass-transfer rate,
and magnetic field that also determine its position among the 
various subtypes of CVs.

In addition, the hibernation model predicts changes of the mass transfer
rate in the evolution of the pre- and post-nova, i.e. a long
state of low mass transfer once the white dwarf has cooled
down and irradiation ceases to push the secondary out of
thermal equilibrium \citep{sharaetal86-1, prialnik+shara86-1}.
It was originally invoked to explain the missing CV population
at minimum orbital period. However, recent surveys like SDSS and CSS
have shown that this is an observational bias, so in principle,
hibernation is no longer needed.
Still, the theoretical arguments for the occurrence of hibernation 
are valid and it thus seems likely that some kind of hibernation occurs.
Unambiguous observational evidence, however, is still missing.

Some new input to the discussion on old novae with low mass transfer rates
came from the discovery of nova shells around Z\,Cam \citep{sharaetal07-1} 
and AT\,Cnc \citep{sharaetal12-1}, both
CVs that were 
known as dwarf novae with no nova outburst recorded. While the existence 
of such systems could be taken as evidence for hibernation, 
it could also be interpreted in the way that all type of CVs (including
low mass transfer dwarf novae) can experience the one or other
nova explosion during their lifetime without necessarily undergoing cyclic
changes of CV-class.

Several years ago, we conducted a project investigating 
old novae which had experienced large outburst
amplitudes \citep{schmidtobreicketal05-1, schmidtobreicketal03-2}. The idea behind this was that since the
absolute magnitude of a nova explosion depends mainly on the mass of the 
white dwarf \citep{livio92-1} and thus differs only slightly for different
systems, novae with large outburst amplitudes are
intrinsically faint CVs and therefore likely candidates for low mass transfer
systems.  Spectroscopic observations seemed to confirm the low mass transfer
status of two of the systems from our sample: V842\,Cen and XX\,Tau. A recent
approach by \cite{tappertetal12-1} to re-discover lost old novae revealed
two more candidates: V2109\,Oph and V728\,Sco. In particular V728\,Sco
which has also been observed photometrically can be considered a low mass
transfer system which shows frequent stunted dwarf novae outbursts 
\cite[and also this conference]{tappertetal13-1}. The various attempts to
determine the orbital periods
of these low mass transfer candidates are listed in Table~\ref{periods_tab}.

\begin{table}[bt]
\caption{\label{periods_tab} The four low mass transfer candidates among the 
old novae and possible values for their orbital period.}
\centerline{
\begin{tabular}{l c l}
\hline\\[-0.35cm]
System & $P_{orb}$ [h] & comments\\[0.05cm]
\hline\\[-0.35cm]
V842\, Cen  & 3.94 or 3.79 & two possible values from high-speed photometry \\
            &              & \citep{woudtetal09-1}\\
            & 3.51 & periodicity in x-ray photometry \citep{lunaetal12-1}\\
V2109\, Oph & - & \\
V728\,Sco   & 3.32 & eclipse measurements \citep{tappertetal13-1} \\
XX\,Tau     & 3.26 & orbital or superhump period from time-resolved\\
            &     &   photometry \citep{rodriguez-gil+torres05-1} \\[0.05cm]
\hline
\end{tabular}}
\end{table}

\section{Combining the information of SW\,Sex and old novae}
\begin{figure}
\resizebox{!}{9.0cm}{\includegraphics{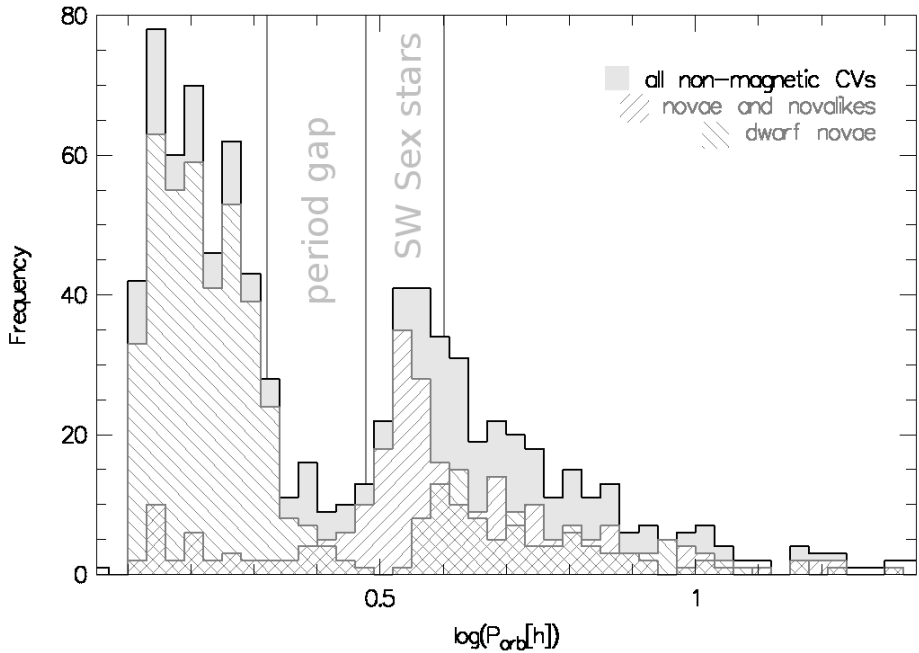}}
 \caption{\label{periods_all} 
The orbital period distribution of the non or weakly magnetic CVs.
Data taken from \cite{ritter+kolb03-1} (update RKcat7.18, 2012).}
\end{figure}
In Fig.~\ref{periods_all}, the orbital period distribution of CVs
is given for different classes of non- or weakly magnetic CVs as defined 
in \cite{ritter+kolb03-1} (update RKcat7.18, 2012). 
The position of the period gap and the location of the SW\,Sex regime 
are indicated.
The lack of dwarf novae
below 3.8\,h and the peak of high mass transfer systems in the same area 
are evident.

In this context, it appears interesting that the possible orbital 
periods of the three low mass transfer old novae with confirmed 
periodicities (see Table~\ref{periods_tab})
seem to lie in the 2.8-4\,h period range which is the regime of the SW\,Sex 
stars. It is not particularly surprising to find the majority of the old novae
clustered in this period range since due to the high mass transfer rate of 
the CVs in this range their recurrence time is lower and therefore they 
are more likely
to be observed as a novae \citep{ibenetal92-1}. However, these three old novae are observed as 
low mass transfer systems! 

The observations therefore seem to indicate that 
these three systems are indeed SW\,Sex stars which due to the nova explosion in 
the recent past were pushed into a low state and thus experience low mass
transfer rates at the moment. This scenario is supported by the likeliness
of the eclipse light curve of V728\,Sco during an outburst with the
eclipse lightcurves of SW\,Sex stars as was pointed out by C. Knigge during
the discussion of C. Tappert's talk of this conference. 

\section{Conclusions}
The finding of
several old novae with currently low mass transfers in a period range that is 
generally populated by high mass transfer systems is interpreted as evidence
that the nova eruption has some effect on the mass transfer rate
similar to what Shara et al. 
proposed for the hibernation scenario. 
Whether the mass transfer in these hibernating systems will completely 
cease or just
remains on a low level before rising up again remains to be investigated. 

In any case, the few low mass transfer systems in the  2.8-4\,h period range
(see Fig.~\ref{periods_all}) that were not observed as novae but are classified 
as dwarf novae, seem to be good candidates to search for indications of 
nova eruptions in the past, i.e. shells as observed around Z\,Cam and
AT\,Cnc.

\acknowledgements This research was supported by FONDECYT Regular grant
1120338 (CT). We gratefully acknowledge the use of 
the SIMBAD database, operated at CDS, Strasbourg, France,
and of NASA's Astrophysics Data System Bibliographic Services.

%\bibliographystyle{asp2010}
%\bibliography{aamnem99,aabib}

\end{document}